\documentclass[
aps,epsf,
superscriptaddress,
twocolumn,
amsmath,amssymbol,
]
{revtex4-1}

\usepackage[dvipdfmx]{graphicx}
\usepackage[dvipdfmx]{color}

\begin{document}


\title{Development of Ferromagnetic Fluctuations in Heavily Overdoped (Bi,Pb)$_2$Sr$_2$CuO$_{6+\delta}$ Copper Oxides}

\author{Koshi Kurashima}
\affiliation{Department of Applied Physics, Tohoku University, 6-6-05 Aoba, Aramaki, Sendai 980-8579, Japan}
\author{Tadashi Adachi}
\email[Corresponding author: ]{t-adachi@sophia.ac.jp}
\affiliation{Department of Engineering and Applied Sciences, Sophia University, 7-1 Kioi-cho, Chiyoda-ku, Tokyo 102-8554, Japan.}
\author{Kensuke M. Suzuki}
\affiliation{Institute for Materials Research, Tohoku University, 2-1-1 Katahira, Sendai 980-8577, Japan}
\author{Yasushi Fukunaga}
\affiliation{Department of Applied Physics, Tohoku University, 6-6-05 Aoba, Aramaki, Sendai 980-8579, Japan}
\author{Takayuki Kawamata}
\affiliation{Department of Applied Physics, Tohoku University, 6-6-05 Aoba, Aramaki, Sendai 980-8579, Japan}
\author{Takashi Noji}
\affiliation{Department of Applied Physics, Tohoku University, 6-6-05 Aoba, Aramaki, Sendai 980-8579, Japan}
\author{Hitoshi Miyasaka}
\affiliation{Institute for Materials Research, Tohoku University, 2-1-1 Katahira, Sendai 980-8577, Japan}
\affiliation{Department of Chemistry, Tohoku University, 6-3 Aoba, Aramaki, Sendai 980-8578, Japan}
\author{Isao Watanabe}
\affiliation{Meson Science Laboratory, Nishina Center for Accelerator-Based Science, RIKEN, 2-1 Hirosawa, Wako 351-0198, Japan}
\author{Masanori Miyazaki}
\affiliation{Graduate School of Engineering, Muroran Institute of Technology, 27-1 Mizumoto, Muroran 050-8585, Japan}
\author{Akihiro Koda}
\affiliation{Institute of Materials Structure Science, High Energy Accelerator Research Organization (KEK-IMSS), 1-1 Oho, Tsukuba 305-0801, Japan}
\author{Ryosuke Kadono}
\affiliation{Institute of Materials Structure Science, High Energy Accelerator Research Organization (KEK-IMSS), 1-1 Oho, Tsukuba 305-0801, Japan}
\author{Yoji Koike}
\affiliation{Department of Applied Physics, Tohoku University, 6-6-05 Aoba, Aramaki, Sendai 980-8579, Japan}

\date{\today}

\begin{abstract}
We demonstrate the presence of ferromagnetic (FM) fluctuations in the superconducting and non-superconducting heavily overdoped regimes of high-temperature superconducting copper oxides, using (Bi,Pb)$_2$Sr$_2$CuO$_{6+\delta}$ (Bi-2201) single crystals.
Magnetization curves exhibit a tendency to be saturated in high magnetic fields at low temperatures in the heavily overdoped crystals, which is probably a precursor phenomenon of a FM transition at a lower temperature.
Muon spin relaxation detects the enhancement of spin fluctuations at high temperatures below 200 K.
Correspondingly, the ab-plane resistivity follows a 4/3 power law in a wide temperature range, which is characteristic of metals with two-dimensional FM fluctuations due to itinerant electrons.
As the Wilson ratio evidences the enhancement of spin fluctuations with hole doping in the heavily overdoped regime, it is concluded that two-dimensional FM fluctuations reside in the heavily overdoped Bi-2201 cuprates, which is probably related to the decrease in the superconducting transition temperature in the heavily overdoped cuprates.
\end{abstract}

\maketitle

In hole-doped high-temperature superconducting cuprates, the relationship between the antiferromagnetism and superconductivity has intensively been studied. 
Antiferromagnetic (AF) fluctuations by which electron paring is believed to be mediated have been observed in the underdoped and optimally doped regimes \cite{Birgeneau2006}.
In the overdoped regime where the superconducting transition temperature ${\textit{T}}_{\rm c}$ is depressed with hole doping, inelastic neutron-scattering \cite{Wakimoto2004} and muon spin relaxation ($\mu$SR) \cite{Risdiana2008} experiments have revealed the weakening of the low-energy AF spin correlation with hole doping.
A recent resonant inelastic x-ray scattering experiment, on the other hand, has revealed that high-energy AF fluctuations persist to the non-superconducting heavily overdoped regime \cite{Dean2013}.
This suggests that the suppression of superconductivity in the heavily overdoped regime might not be related to AF fluctuations.

In the heavily overdoped regime, unlike the general belief of the nonmagnetic Fermi-liquid-like ground state, phenomena incompatible with a simple Fermi-liquid picture have been observed.
The Curie constant has increased with hole doping in overdoped Tl${_2}$Ba${_2}$CuO$_{6+{\rm \delta}}$ (Tl-2201) \cite{Kubo1991}, La${_{2-x}}$Sr${_x}$CuO$_{4}$ (LSCO) and La${_{2-x}}$Ba${_x}$CuO$_{4}$ \cite{Oda1991}.
The ab-plane electrical resistivity $\rho_{\rm ab}$ has not exhibited a $T^2$ behavior in heavily overdoped LSCO \cite{Nakamae2003}.
Therefore, there might exist other ordered states hidden adjacent to the superconducting phase.

Kopp \textit{et al.} have insisted in terms of the quantum critical scaling theory that the non-Fermi-liquid-like temperature dependence of the magnetic susceptibility $\chi$ in non-superconducting heavily overdoped Tl-2201 is due to the existence of a ferromagnetic (FM) phase \cite{Kopp2007}.
Electronic band calculations have suggested that the ferromagnetism appears locally around Ba clusters in overdoped $\rm La_{2-\it x\rm}Ba_{\it x\rm}CuO_4$ \cite{Barbiellini2008}.
A recent theoretical calculation of the spin dynamical structure factor by the determinant quantum Monte Carlo method has supported the occurrence of the ferromagnetism in the heavily overdoped regime \cite{Jia2014}.
Experimentally, Sonier \textit{et al.} have reported from zero-field $\rm \mu$SR measurements in non-superconducting heavily overdoped LSCO that the relaxation rate of muon spins is enhanced with decreasing temperature below 0.9 K, 
suggesting the development of spin fluctuations \cite{Sonier2010}.
They have also reported that $\rho_{\rm ab}$ exhibits a $T^{5/3}$ behavior in a wide temperature range from 60 K to room temperature.
The $T^{5/3}$ behavior is characteristic of metals with three-dimensional FM fluctuations due to itinerant electrons, according to the self-consistent renormalization (SCR) theory
of spin fluctuations \cite{Ueda1975}.
The $T^{5/3}$ behavior has also been observed in the itinerant-electron ferromagnet $\rm Y_4Co_3$ above the Curie temperature \cite{Kolodziejczyk1984}.
These suggest that three-dimensional FM fluctuations exist in non-superconducting heavily overdoped LSCO.
The FM fluctuations may be related to the decrease in ${\textit{T}}_{\rm c}$ with hole doping in the overdoped regime owing to the competition between AF and FM fluctuations.
However, since effects of disorder due to the dopant Sr and/or oxygen deficiency in the $\rm CuO_2$ plane might be related to the FM fluctuations in LSCO, the universality of the FM fluctuations among cuprates is unclear. 

In this Letter, we study the electronic states in heavily overdoped $\rm (Bi,Pb)_2Sr_2CuO_{6+\delta}$ (Bi-2201) cuprates.
In Bi-2201, we control the hole concentration by changing the amount of excess oxygen $\rm \delta$, indicating no substitution-induced disorder and no oxygen deficiency in the $\rm CuO_2$ plane.
Static magnetization curves showed a tendency to be saturated in high magnetic fields at low temperatures, which is probably a precursor phenomenon of a FM transition at a lower temperature.
The development of spin fluctuations with decreasing temperature as well as with increasing hole concentration was also observed in zero-field $\rm \mu$SR measurements.
The $\rho_{ab}$ was almost proportional to $T^{4/3}$ over a wide temperature range, which is characteristic of two-dimensional FM fluctuations due to itinerant electrons \cite{Hatatani1995}.
The Wilson ratio increased with hole doping, suggesting the enhancement of spin fluctuations.
Accordingly, two-dimensional FM fluctuations due to itinerant electrons probably exist in heavily overdoped Bi-2201.

Crystals obtained from four grown rods \cite{Chong1997} were used in the following results; Bi-2201[A] - $\rm Bi_{1.76}Pb_{0.35}Sr_{1.89}CuO_{6+\delta}$, Bi-2201[B] - $\rm Bi_{1.71}Pb_{0.32}Sr_{1.97}CuO_{6+\delta}$, Bi-2201[C] - $\rm Bi_{1.77}Pb_{0.33}Sr_{1.90}CuO_{6+\delta}$, Bi-2201[D] - $\rm Bi_{1.76}Pb_{0.35}Sr_{1.89}CuO_{6+\delta}$.
The hole concentration per Cu $p$ was determined by the empirical equation and the thermoelectric power \cite{Obertelli1992,Presland1991,Kudo2009}, described in the Supplemental Material, although how to estimate the $p$ value is an unsettled issue in the Bi-2201 cuprates \cite{Kudo2009,AndoPRB,Kondo}.
Figure 1(a) shows the temperature dependence of the static $\chi$ of Bi-2201[A].
The inset shows the $T_{\rm c}$ vs $p$ plot.
For the overdoped crystals with $p \leq 0.232$, $\chi$ is almost independent of temperature in the normal state above 15 K, while $\chi$ for the heavily overdoped crystals with $p \geq 0.257$ rapidly increases with decreasing temperature.
The increase in $\chi$ has also been observed in heavily overdoped LSCO \cite{Takagi1989} and Tl-2201 \cite{Kubo1991}, and therefore it is reminiscent of FM fluctuations.
In fact, $\chi$ is proportional to $T^{-5/4}$ as shown in Fig. 1(b), which is consistent with the quantum critical scaling theory of FM fluctuations \cite{Kopp2007}.

\begin{figure*}[tbp]
\begin{center}
\includegraphics[width=0.7\linewidth]{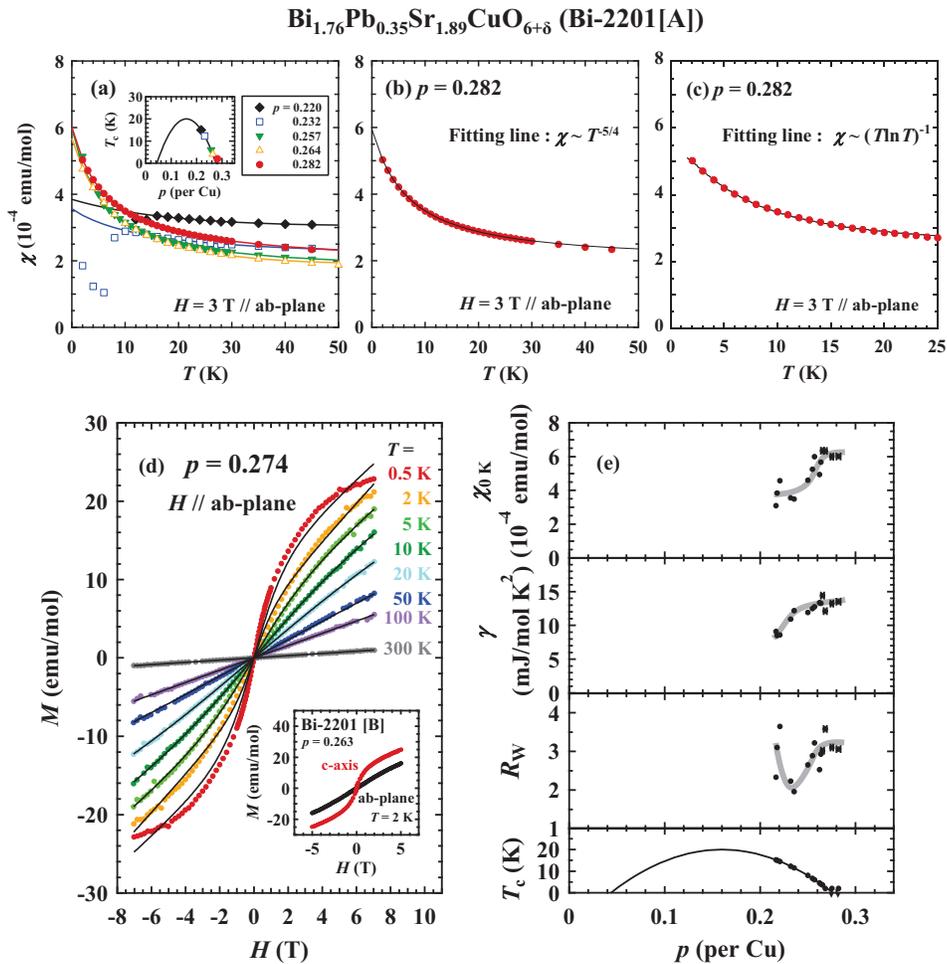}
\caption{
(a) Temperature dependence of the magnetic susceptibility $\chi$ in overdoped and heavily overdoped Bi-2201[A] in a magnetic field of 3 T along the ab-plane. 
Solid lines are fitting results using Eq. (2) in the text. 
The inset shows the \textit{T}$_{\rm c}$ vs \textit{p} plot. 
(b)(c) Temperature dependence of $\chi$ of heavily overdoped Bi-2201[A] with \textit{p} = 0.282. 
The solid line in (b) indicates the fitting result using $\chi$ = $\chi _{\rm 0\it}$ + \textit{C}/(\textit{T} - \textit{T}$_{\rm 0\it}$)$^{5/4}$ and that in (c) indicates the fitting result using $\chi$ = $\chi _{\rm 0\it}$ + \textit{C}/(\textit{T}ln\textit{T} - \textit{T}$_{\rm 0\it}$).
(d) Magnetization curves of heavily overdoped Bi-2201[A] in magnetic fields along the ab-plane. 
Solid lines are fitting results using the standard paramagnetic Brillouin function. 
(Inset) Magnetization curves of heavily overdoped Bi-2201[B] in magnetic fields along the ab-plane and c-axis at 2 K. 
(e) Hole-concentration dependence of the extrapolated value of $\chi$ at 0 K $\chi_{0\rm K}$, the electronic specific heat coefficient $\gamma$, the Wilson ratio \textit{R}$_{\rm W}$ and \textit{T}$_{\rm c}$ in Bi-2201[A].
}  
\label{fig:Figure1}
\end{center}
\end{figure*}

Pronounced features of FM fluctuations are seen in the static magnetization curve of heavily overdoped Bi-2201[A] in Fig. 1(d).
It is found that the magnetization curves are linear in $H$ above 20 K, while they tend to be saturated in high magnetic-field regions below 20 K.
These behaviors cannot be reproduced in terms of the standard paramagnetic Brillouin function as shown in Fig. 1(d).
It is noted that the nonlinear magnetization curve is highly anisotropic as displayed in the inset, eliminating magnetic impurities as an origin of the nonlinear behavior.
Considering no hysteresis in the magnetization curve down to 0.5 K, the saturated magnetization in high magnetic fields at low temperatures is probably a precursor phenomenon of the FM transition at a lower temperature.

\begin{figure}[tbp]
\begin{center}
\includegraphics[width=1.0\linewidth]{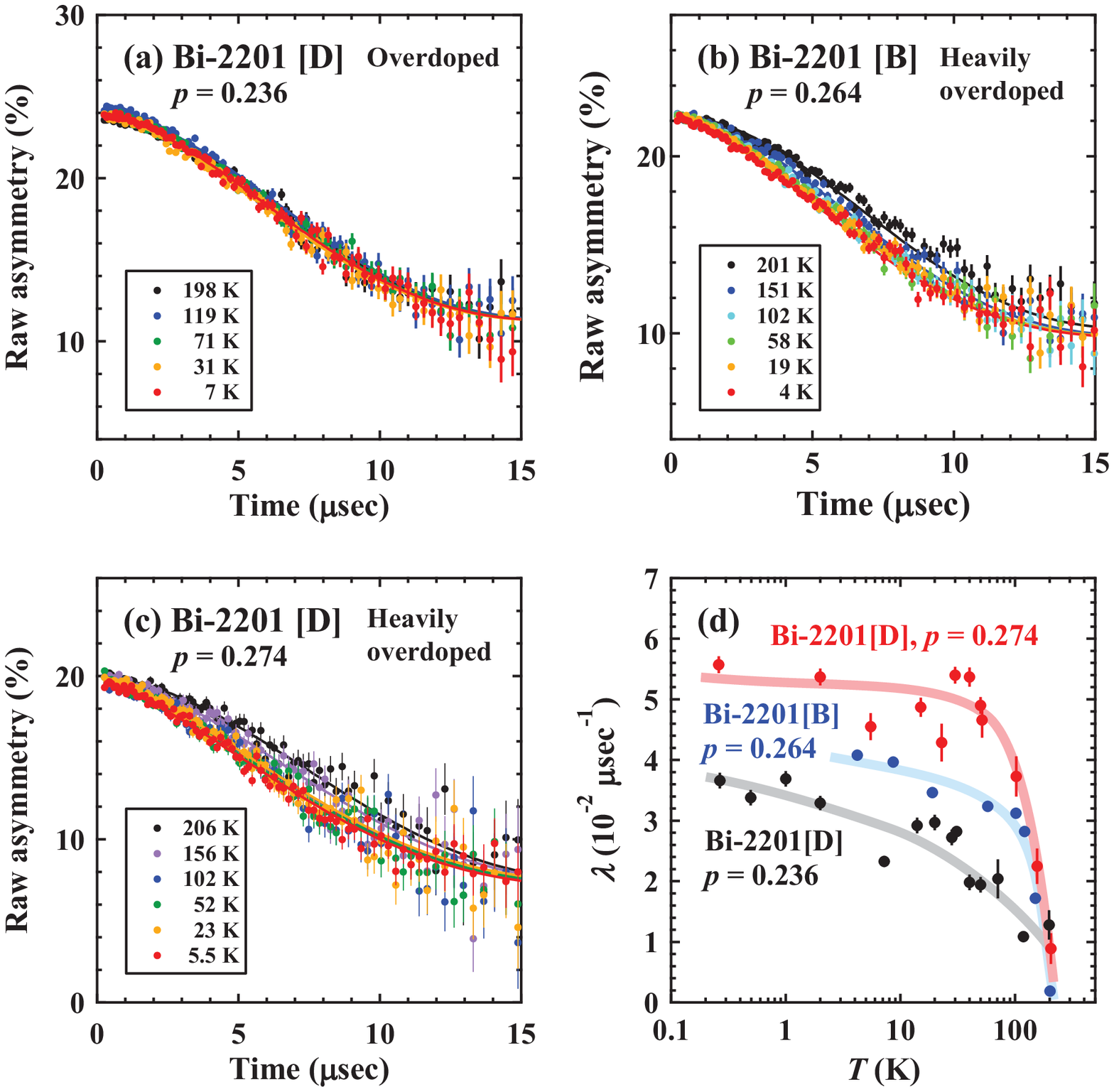}
\caption{
Zero-field $\rm \mu$SR time spectra of (a) overdoped Bi-2201[D], (b) heavily overdoped Bi-2201[B], (c) heavily overdoped Bi-2201[D]. 
Solid lines are best-fit results using Eq. (1) in the text. 
(d) Temperature dependence of the relaxation rate of muon spins $\lambda $. 
Error bars represent fitting errors using Eq. (1).
Solid lines are to guide the reader's eye.
}
\label{fig:Figure2}
\end{center}
\end{figure}

To investigate dynamic behaviors of spins, zero-field $\rm \mu$SR measurements were performed in Bi-2201[B] and [D] shown in Figs. 2(a-c).
While the zero-field $\rm \mu$SR time spectra are more or less Gaussian-like originating from nuclear-dipole fields, the spectra of heavily overdoped Bi-2201[B] and [D] show the enhancement of the relaxation of muon spins with decreasing temperature.
These are suggestive of the development of spin fluctuations.
At low temperatures below 50 K down to 0.3 K, the change of the spectra becomes small as shown in Fig. S1 of the Supplemental Material.
The spectra are fitted using the following function:
\begin{equation}
A(t) = A_0 {\rm exp} (-\lambda t)\, G_{\rm z} (\Delta ,t) + A_{\rm BG}.
\end{equation}
The $A_0$ is the initial asymmetry, $\lambda $ is the relaxation rate of muon spins, corresponding to the development of spin fluctuations, $G_{\rm z}(\Delta,t)$ is the static Kubo-Toyabe function \cite{Kubo1967}, $\it\Delta $ the distribution width of the nuclear-dipole field at the muon site, and $A_{\rm BG}$ is the temperature-independent background term.
In the analysis, $\Delta$ was treated as temperature independent. (see the Supplemental Material.) 
In Fig. 2(d), we show the temperature dependence of $\lambda $. 
It is clearly seen that $\lambda $ increases with decreasing temperature below 200 K, and the increase becomes larger with hole doping. 
These suggest that spin fluctuations, which is probably FM, are developed at high temperatures and are enhanced with hole doping in heavily overdoped Bi-2201.

Transport and thermal properties of heavily overdoped Bi-2201 exhibit temperature dependences characteristic of FM fluctuations.
Figure 3(a) shows the temperature dependence of $\rho _{\rm ab}$ in Bi-2201[C].
The $\rho _{\rm ab}$ exhibits a $T^{4/3}$ behavior over a wide temperature range for $p \geq 0.271$, as reported in overdoped $\rm Bi_2Sr_{1.6}La_{0.4}CuO_{\it y}$ \cite{Konstantinovic2001}. 
Figure 3(b) shows the $p$ dependence of the exponent $n$ obtained by expressing $\rho _{\rm ab}$ in the normal state as $\rho _{\rm ab} = \it \rho _{ \rm 0\it } + AT^n$, where $\it \rho _{ \rm 0\it }$ is the temperature-independent term and $A$ is the temperature coefficient.
Although the data exhibit a spread to some extent, it is found that $n \sim 1$ near the optimally doped regime and that $n$ increases with underdoping, which is consistent with preceding results \cite{Ando2004}. (The resistivity data are shown in Fig. S2 of the Supplemental Material.)
With overdoping, $n$ increases and converges to 4/3 in the heavily overdoped regime \cite{Kurashima2014}, which is incompatible with the nonmagnetic Fermi-liquid-like $T^2$ behavior.
According to the SCR theory, the $T^{4/3}$ behavior of $\rho_{\rm ab}$ is characteristic of metals with two-dimensional FM fluctuations due to itinerant electrons \cite{Hatatani1995} as observed in $\rm Sr_{2-\it y\rm}La_{\it y\rm}RuO_4$ \cite{Kikugawa2004}, as well as $\rm Sr_2RuO_4$ \cite{Yoshida1998} and $\rm CePt$ \cite{Larrea2005} under pressures.

The inset of Fig. 3(c) shows the temperature dependence of the specific heat of Bi-2201[A] plotted as $C/T$ vs. $T^2$ at low temperatures down to 0.4 K.
The heavily overdoped crystal with $p$ = 0.277 exhibits deviation from the linearity and following anomalous upturn below 2 K, which has been observed also in $\rm Sr_{1.8}La_{0.2}RuO_4$ with FM fluctuations \cite{Kikugawa2004}.
In the main figure, we plot the magnetic specific heat $C_m = C - \gamma T - \beta T^3$ of heavily overdoped Bi-2201[D] as $C_m/T$ vs. $T^{-1/3}$, where $\gamma T$ and $\beta T^3$ are the electronic and phonon specific heats, respectively.
Obviously, $C_m/T$ is proportional to $T^{-1/3}$ below 2 K, suggesting the occurrence of two-dimensional FM fluctuations \cite{Hatatani1995}.
It is noted that, as shown in Fig. 1(c), $\chi$ almost follows $(T{\rm ln}T)^{-1}$ at low temperatures, also suggesting the two-dimensional FM fluctuations.

\begin{figure}[tbp]
\begin{center}
\includegraphics[width=1.0\linewidth]{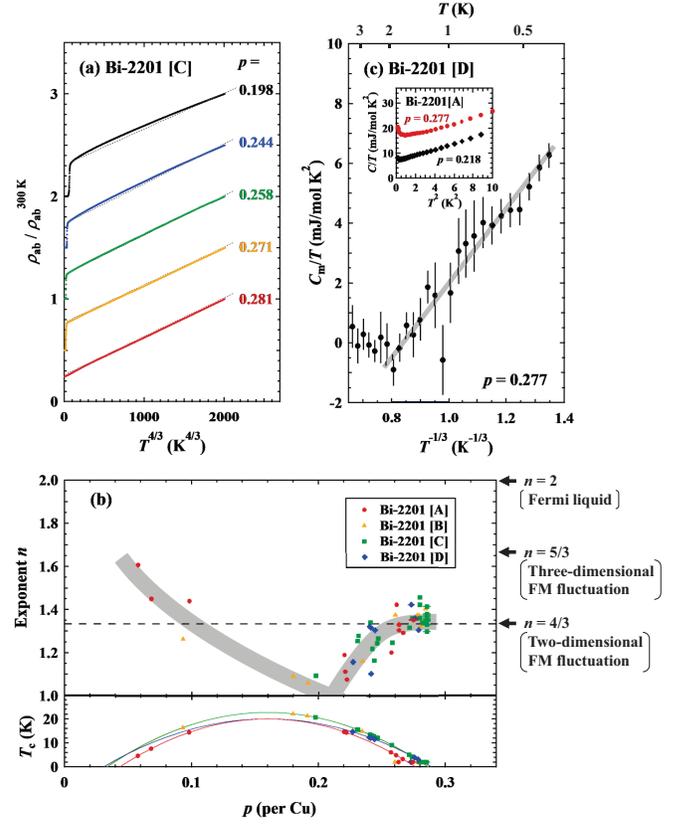}
\caption{
(a) Temperature dependence of the ab-plane electrical resistivity $\rho_{\rm ab}$ normalized by $\rho_{\rm ab}$ at 300 K plotted against \textit{T}$^{4/3}$ in Bi-2201[C]. 
The data are shifted upward by the same offset. 
(b) Hole-concentration dependence of the temperature exponent \textit{n} of $\rho _{\rm ab}$ = $\rho _{\rm 0}$ + \textit{AT}$^n$ and \textit{T}$_{\rm c}$ in Bi-2201[A], [B], [C], [D]. 
The solid line in \textit{n} vs \textit{p} is to guide the reader's eye. 
Solid lines in \textit{T}$_{\rm c}$ vs \textit{p} are the empirical parabolic function of \textit{T}$_{\rm c}$ [K] = 20.0 - 74.02(\textit{p} [per Cu] - 0.16), \textit{T}$_{\rm c}$ [K] = 22.7 - 62.14(\textit{p} [per Cu] - 0.16), \textit{T}$_{\rm c}$ [K] = 22.7 - 62.83(\textit{p} [per Cu] - 0.16) and \textit{T}$_{\rm c}$ [K] = 20.0 - 60.05(\textit{p} [per Cu] - 0.16) for Bi-2201[A], [B], [C] and [D], respectively.
(c) Temperature dependence of the magnetic specific heat \textit{C}$_m$ = \textit{C} - $\gamma$\textit{T} - $\beta$\textit{T}$^3$ plotted as \textit{C}$_m$/\textit{T} vs. \textit{T}$^{-1/3}$ in Bi-2201[D]. 
Error bars represent measurement errors.
(Inset) Total specific heat plotted as \textit{C}/\textit{T} vs. \textit{T}$^2$ in Bi-2201[A].
}
\label{fig:Figure3}
\end{center}
\end{figure}

To obtain further insights into the FM fluctuations, we discuss the Wilson ratio $R_{\rm W}$ relating to spin fluctuations.
To estimate $R_{\rm W}$, the extrapolated value of $\chi$ at 0 K, $\chi_{0 {\rm K}}$, was used from the fitting of $\chi$ between $T_{\rm c}$ and 40 K with the following equation
\begin{equation}
\chi = \chi_{\rm const} + C/(T - T_0) - DT, 
\end{equation}
as shown in Fig. 1(a).
The first term represents the temperature-independent susceptibility, the second represents the Curie-Weiss-like upturn, and the third represents the temperature-dependent Pauli susceptibility due to the thermal modification of the density of states at the Fermi level as reported in Bi-2201 \cite{LeBras2002}.
The value of $\gamma$ was obtained by fitting the specific heat in a magnetic field of 9 T suppressing the superconductivity with $C/T = \gamma + \beta T^2$ in a $T^2$ range of $\rm 4 - 8\, K^2$ as shown in Fig. S3 of the Supplemental Material.
The $p$ dependences of $\chi_{0\rm K}$, $\gamma$ and $R_{\rm W} = \pi^2 k_{\rm B}^2 \chi_{0 {\rm K}} / (3 \mu_{\rm B}^2 \gamma)$ of Bi-2201[A] are displayed in Fig. 1(e), where $k_{\rm B}$ is the Boltzmann constant and $\mu_{\rm B}$ is the Bohr magneton.
Both $\chi_{0 {\rm K}}$ and $\gamma$ increase with $p$ and tend to be saturated in the heavily overdoped regime, suggesting the enhancement of the density of states at the Fermi level.
Surprisingly, $R_{\rm W}$ increases with hole doping in the heavily overdoped regime of $p > 0.23$, suggesting the enhancement of spin fluctuations.
The enhanced $R_{\rm W}$ has also been observed in $\rm Ca_{2-\it x}\rm Sr_{\it x}\rm RuO_4$ around $x = 0.5$ in which FM fluctuations are significant \cite{Nakatsuji2003}.

The present results in heavily overdoped Bi-2201 are summarized as follows.
The $\chi$ exhibits an upturn at low temperatures.
The magnetization tends to be saturated in high magnetic fields at low temperatures.
Zero-field $\rm \mu$SR reveals the development of spin fluctuations below 200 K.
The $\rho_{\rm ab}$ exhibits a $T^{4/3}$ behavior in a wide temperature range.
The magnetic specific heat follows $T^{-1/3}$ dependence.
The $R_{\rm W}$ increases with hole doping.
All these results strongly suggest that there exists two-dimensional FM fluctuations in heavily overdoped Bi-2201. 
For the difference of the dimensionality of FM fluctuations between Bi-2201 and LSCO \cite{Sonier2010}, it is probably originated from the crystal structure.
The distance between the nearest neighboring $\rm CuO_2$ planes in Bi-2201 is approximately twice as long as that in LSCO.
Moreover, the van der Waals bond between atoms is included in the blocking layer of Bi-2201, while the bonding in the blocking layer is ionic in LSCO.
Zero-field $\rm \mu$SR reveals that FM fluctuations are developed below 200 K in Bi-2201, whereas they are developed below 0.9 K in LSCO \cite{Sonier2010}. 
These are also understandable in terms of the dimensionality of the FM fluctuations.
Originating from the structure of cuprates, the two-dimensional FM fluctuations might be developed at high temperatures and FM fluctuations might become three-dimensional at low temperatures.
The two-dimensional FM fluctuations might be developed at high temperatures also in LSCO \cite{LSCO-2D} and the three-dimensional FM fluctuations might be developed below $\sim$ 0.3 K in heavily overdoped Bi-2201.

\begin{figure}[tbp]
\begin{center}
\includegraphics[width=1.0\linewidth]{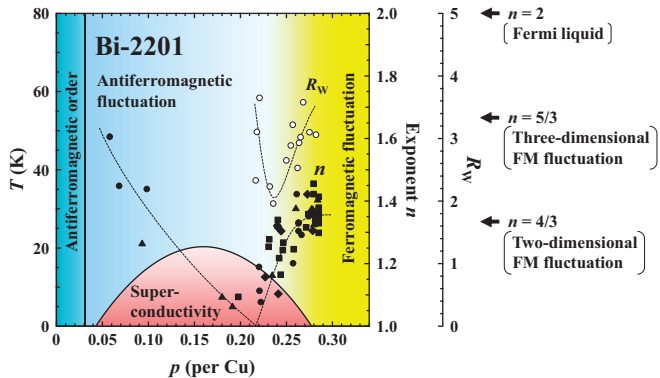}
\caption{
Schematic phase diagram of Bi-2201. 
Open circles represent \textit{R}$_{\rm W}$ of Bi-2201[A]. 
Closed circles, triangles, squares and diamonds represent the exponent \textit{n} of $\rho _{\rm ab} = \it \rho _{ \rm 0\it } + AT^n$ of Bi-2201[A], [B], [C], [D], respectively. 
Dashed lines are to guide the reader's eye.
}
\label{fig:Figure4}
\end{center}
\end{figure}

In Fig. 4, a new phase diagram of Bi-2201 is proposed including the $p$ dependences of $R_{\rm W}$ and $n$. 
As characterized by the increase in $R_{\rm W}$ and the saturation of $n$ around 4/3, the region of FM fluctuations is shown in the heavily overdoped regime.
It is found that the magnetic ground state changes from the AF to FM one with hole doping.
The FM fluctuations exist even in the superconducting heavily overdoped regime, implying the interference between FM fluctuations and the electron paring mediated by AF fluctuations and resulting in the decrease in $T_{\rm c}$ with hole doping in the heavily overdoped regime \cite{Kopp2007}.

There exist two candidates for the origin of FM fluctuations.
One is the metallic ferromagnetism due to enhanced spin fluctuations, the large density of states at the Fermi level and the good Fermi-surface nesting with the nesting vector of \mbox{\boldmath $q$} $\to 0$.
In fact, the value of $R_{\rm W}$ indicates that spin fluctuations are enhanced in the heavily overdoped regime.
It has been reported from the angle-resolved photoemission spectroscopy \citep{Kondo2004,Hashimoto2008} and scanning tunneling spectroscopy \cite{Piriou2011} that the van Hove singularity resides close to the Fermi level in heavily overdoped Bi-2201, suggesting the large density of states at the Fermi level in the heavily overdoped regime.
A theoretical calculation based on the three-band model has suggested that the \mbox{\boldmath $q$} position where the spin susceptibility is enhanced evolves from \mbox{\boldmath $q$} $= (\pi,\pi)$ in the parent compound toward \mbox{\boldmath $q$} $= (0,0)$ with hole doping \cite{Ogura2015}.
This situation is quite similar to that of $\rm Sr_{2-\it y\rm}La_{\it y\rm}RuO_4$ with FM fluctuations \cite{Kikugawa2004}.
All these are consistent with the occurrence of the metallic ferromagnetism.
The other candidate is the double exchange interaction due to the multiband structure.
This is because Compton-scattering measurements in LSCO \cite{Sakurai2011} have suggested that holes are doped mainly into the $\rm Cu\, 3d_{3\it z^{2}-r^{2}\rm}$ orbital in the heavily overdoped regime, producing both $\rm Cu\,  3d_{\it x^{2}-y^{2}\rm}$ and $\rm Cu \, 3d_{3\it z^{2}-r^{2}\rm}$ spins and generating the FM interaction due to the Hund coupling.

In conclusion, FM fluctuations exist in heavily overdoped Bi-2201, suggesting the universal feature of the hole-doped cuprates.
The magnetic ground state changes from the AF to FM one with hole doping. 
Moreover, the FM fluctuations are probably related to the suppression of superconductivity in the heavily overdoped regime. 
The FM fluctuations may answer several unsolved non-Fermi-liquid-like behaviors in the heavily overdoped regime.
For example, the broadening of nodal quasiparticle peaks observed in the angle-resolved photoemission spectroscopy of $\rm La_{1.78}Sr_{0.22}CuO_4$\cite{Zhou2004} and Tl-2201 \cite{Peets2007,Plate2005} may be predominantly caused by the scattering of quasiparticles by low-energy FM fluctuations \cite{Kopp2007}.
The more detailed relationship between the FM fluctuations and superconductivity in cuprates should be clarified in future.

We thank A. Sakuma, Y. Shimizu, H. Tsuchiura, K. Ueda, H. Yokoyama for informative discussions. 
We are indebted to M. Ishikuro for his help in the ICP analysis. 
$\rm \mu$SR measurements were partially supported by MEXT KAKENHI Grant Number 23108004 and by the KEK-MSL Inter-University Research Program (Proposals No. 2015A0198 and No. 2016B0121). 
K. K. thanks the Tohoku University Division for Interdisciplinary Advanced Research and Education for their financial support.



\begin{thebibliography}{99}
\bibitem{LSCO-2D}
The two-dimensional FM fluctuations have not yet been observed in the heavily overdoped LSCO due to lack of zero-field $\mu$SR results at high temperatures.
Neutron-scattering experiments detecting fluctuations with higher frequencies than $\mu$SR, focusing on the FM fluctuations, have not yet been reported in the heavily overdoped LSCO, to our knowledge.
\end{thebibliography}
\end{document}